\documentclass[review]{elsarticle}
\usepackage{lineno,hyperref}
\usepackage[english]{babel}
\usepackage[numbers]{natbib}
\usepackage{xcolor}
\usepackage{latexsym,amsmath,amssymb,amsbsy,graphicx,geometry}
\modulolinenumbers[5]

\begin{document}
	
\begin{frontmatter}
	
	\title{Physical nature of quasi-stable structures existing in antimony melt}
	
	\author[kfu,urfc]{Artem A. Tsygankov\corref{cor1}} 
	\cortext[cor1]{Corresponding author}
	\ead{tsigankov.artiom@yandex.ru}
	
	\author[kfu,ufrc]{Bulat N. Galimzyanov}
	\ead{bulatgnmail@gmail.com}
	
	\author[kfu,ufrc]{Anatolii V. Mokshin}
	\ead{anatolii.mokshin@mail.ru}
	
	\affiliation[kfu]{organization={Kazan Federal University},
		addressline={Kremlevskaya 16},
		city={Kazan},
		postcode={420008},
		state={Tatarstan Republic},
		country={Russia}}
	\affiliation[ufrc]{organization={Udmurt Federal Research Center of the Ural Branch of RAS},
		addressline={Tatyana Baramzina 34}, 
		city={Izhevsk},
		postcode={426067}, 
		state={Udmurtia Republic},
		country={Russia}}
	
	\begin{abstract}
		Equilibrium antimony melt near the melting temperature is characterised by structural features that are not present in simple single-component liquids. The cause of these features may be long-lived structural formations that are not yet fully understood. The present work provides the detailed characterization of the structures formed in liquid antimony near the melting temperature based on the results of quantum chemical calculations and the available neutron and X-ray diffraction data. The quasi-stable structures in antimony melt are detected with lifetimes exceeding the structural relaxation time of this melt. These structures are characterised by a low degree of order and spatial localisation. It is shown for the first time that the elementary units of these quasi-stable structures are triplets of atoms with characteristic lengths of $3.07$\,\AA~and $4.7$\,\AA~and characteristic angles of $45$ and $90$ degrees. It was found that these triplets can form chains and percolating clusters up to $\sim15$\,\AA~in length. The characteristic lengths of these triplets are fully consistent with the correlation lengths associated with short-range order in the antimony melt as determined by diffraction experiments.
	\end{abstract}
	
	
\end{frontmatter}

\section{Introduction}

Antimony is a metalloid with a wide range of practical applications. In particular, antimony and its compounds are used in the manufacture of lithium-ion battery components~\cite{Alcanfor_2024_sb, Wang_2020_sb_application}, catalysts and stabilizers in the production of polyethylene triftonate~\cite{Filella_2020_sb_appication}, adsorbents~\cite{Nagarajan_2020_sb}, in solar cells~\cite{Thirumal_2022_sb_appl} and in memory devices~\cite{Le_Gallo_2020_sb_pcm,Zhou_2013_sb_pcm,Burr_2016_pcm_overview}. A correct understanding the structural characteristics of the melt is important in the production of antimony containing materials. The structure of the initial liquid raw material can determine the physical and mechanical properties of the final product~\cite{Le_Gallo_2020_sb_pcm,	 Dembele_2022_sb_application,Dupont_2016_sb_application,Periferakis_2022}.

Antimony melt has a complex structure different from simple liquids. In liquid antimony, there are the structural features in the form of a pronounced shoulder in the right wing relative to the main maximum of the static structure factor $S(k)$ and of the radial distribution function $g(r)$. These features are detected in neutron and X-ray diffraction experiments~\cite{Jones_2017_sb_aimd,Waseda_1971_experimental_data}. The cause of such features may be the presence of random packing atoms (as it was shown in the work~\cite{Hafner_1992}), the peculiarity of the interatomic interaction, which is anisotropic in nature~\cite{Bichara_1993}, the possible presence of residual crystalline inclusions~\cite{Hao_2013_aimd}, and the presence of clusters and dimers as revealed by ab-initio molecular dynamics simulations~\cite{Jones_2017_sb_aimd,Shi_2024_in_sn_sb}. In addition, the results of recent studies reveal low density regions in liquid antimony. Their formation may be associated with short and medium-range order features that are atypical for metal melts~\cite{Jones_2017_sb_aimd,Akola_2014}. It is noteworthy that such structural features are also observed in melts of other metals belonging to elements of different subgroups, such as zinc, boron, carbon and all the pnictogens~\cite{Jank_1990_group4overview,Jank_1991_group1overview, Jank_1990_group2overview, Hafner_1990_group3overview,Mokshin_Galimzyanov_2018, Mayo_2013_trends}. The existence of anomalies in the structure of these melts can be explained by the presence of the so-called extended short-range order, as in the case of gallium~\cite{Mokshin_2020_gallium}, or by the presence of quasi-stable structures formed by several atoms, as those discovered in liquid bismuth near the melting temperature~\cite{Orton_1975_model_liquid_ge,Orton_1979_liquid_bi_sn,Galimzyanov_2023_liquid_bi}.

Previous studies have examined the structural properties of liquid antimony from a different points of view. For example, the antimony melt is considered to be a bidisperse system near the melting temperature~\cite{Orton_1976_model_liquid_sb}. This approach approximates the structure of liquid antimony in such a way that the structure factor can be decomposed into two independent components corresponding to different sets of atomic diameters. The anomalous properties of liquid antimony were studied by Hafner and Jank~\cite{Hafner_1990_group3overview}. They found that the structural anomaly in antimony melt with the interaction approximation as a pair potential appears at wave number $\sim2k_f$, where $k_f$ is the modulus of the Fermi vector~\cite{Silbert_1976}. However, there is currently no uniform interpretation of anomalies in the structure of liquid antimony.

The aim of the present work is to study in detail the microscopic structure of liquid antimony near its melting temperature and to interpret the observed structural anomalies. The research will be based on quantum chemical calculations and on available X-ray and neutron diffraction experiments. 

\section{Antimony phase diagram}

\begin{figure}[ht!]
	\centering
	\includegraphics[width=1\linewidth]{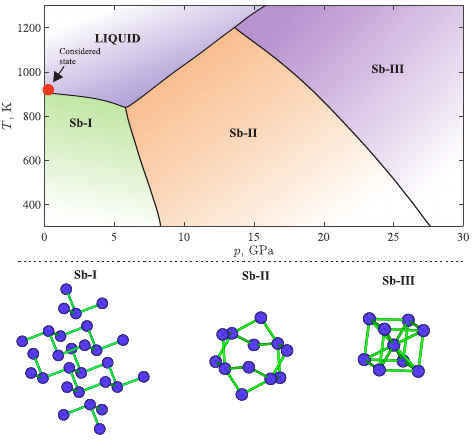}
	\caption{Phase diagram of antimony constructed on the basis of data from Ref.~\cite{Coleman_2018_sb_phase_diagram}. The liquid state with the temperature $T=923$\,K and the pressure $p=1$\,atm considered in the given study is marked by the red filled circle.}
	\label{sb_phase_diagram}
\end{figure}

It is known that in the pressure range $p\in[0,\,35]$~GPa and at temperatures up to $T\approx1300$~K, the phase diagram of antimony has three different crystalline phases -- Sb-I, Sb-II and Sb-III (see Figure~\ref{sb_phase_diagram}). A feature of the phase diagram is that the melting temperature decreases slightly from $T_{m}\approx900$~K to $T_{m}\approx850$~K with increasing pressure to $p\approx7$\,GPa. This negative slope of the melting line $T_m(p)$ is due to the fact that at the phase transition from a crystalline to a liquid phase is the density of a melt is increased~\cite{Jones_2017_sb_aimd}. At pressures $p>7$\,GPa the melting line has a positive slope, where the melting temperature increases with increasing pressure.

The crystalline phase Sb-I is stable under normal conditions and is characterised by the rhombohedral structure with the unit cell parameters $a=b=4.35$~\AA, $c=11.49$~\AA, $\alpha = \beta = 90^\circ$, $\gamma = 120^\circ$ with $12$ atoms and the symmetry group $R\overline{3}m$. Here $a$, $b$, $c$ are the lengths of the edges; $\alpha$, $\beta$, $\gamma$ are the angles between the edges of the unit cell. This crystalline phase has a layered structure where the distance between the layers is $\approx3.36$~\AA. The characteristic lengths of the interatomic bonds are $\approx2.91$~\AA~within each layer. At pressures above $p\simeq9$~GPa there are high density crystalline antimony phases with tetragonal and cubic lattices. In the pressure range $p\in[9;\,27]$~GPa and temperatures $T<1200$\,K the tetragonal Sb-II phase is formed with the unit cell parameters $a=b=8.06$~\AA, $c=4.05$~\AA~ and $\alpha = \beta = \gamma = 90^\circ$ (the symmetry group of the cell is $I4/mcm$ with $9$ atoms). The Sb-III phase has a cubic lattice and is realised at pressures $p>15$\,GPa. The unit cell parameters of this phase are $a=b=c=3.74$~\AA~and $\alpha = \beta = \gamma = 90^\circ$. This high temperature phase also occurs at normal temperatures, where the pressure exceeds $27$~GPa. The unit cells for each type of the antimony crystal lattice are shown in Figure~\ref{sb_phase_diagram}.

In the present work we consider an equilibrium antimony melt with the temperature $T=923$\,K at the pressure $p=1$~atm ($\sim1.0\times10^{-4}$\,GPa). On this isobar, the liquid phase of antimony covers the temperature range $T\in[904;\,1908]$\,K. The melting temperature is $T_m\simeq903.9$~K, while the boiling temperature is $T_b\simeq 1908$~K~\cite{Kim_2023_pubchem_database}.

\section{Structure of liquid antimony}

\begin{figure}[h!]
	\centering
	\includegraphics[width=1.0\linewidth]{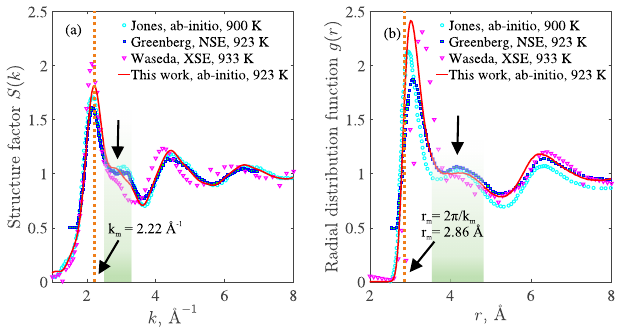}
	\caption{(a) Static structure factor $S(k)$ and (b) radial distribution function $g(r)$ of liquid antimony. The \textit{ab-initio} molecular dynamics simulation results at the tempperature $900$~K, X-ray (XSE) and neutron (NSE) diffraction data are taken from Refs.~\cite{Jones_2017_sb_aimd,Waseda_1971_experimental_data,Greenberg_2010_sb_experiment}.}
	\label{gr_and_sk}
\end{figure}

Figure~\ref{gr_and_sk} shows the experimentally detected structural features in antimony melt, which appear near the melting temperature in the form of a shoulder on the right wing of the main maximum of the static structure factor $S(k)$~\cite{Waseda_1971_experimental_data}. The shoulder in the static structure factor is located in the wave number range $k=[2.7;~3.1]$~\AA$^{-1}$. These structural features also appear in the radial distribution function $g(r)$ as an additional peak at the distance interval $r=[3.8;~4.4]$~\AA~[see Figure~\ref{gr_and_sk}(b)]. The radial distribution function $g(r)$ has an additional peak at the distance interval $r=[3.8;~4.4]$~\AA. In neutron (NSE) diffraction experiments, the shoulder and additional peak appear more pronounced than in X-ray (XSE) diffraction data~\cite{Greenberg_2010_sb_experiment}. This may be mainly due to the existence of a relationship between the observed structural anomaly and the properties of the atom, such as the structure of the electron shell~\cite{Reimers_2015_bond_angle}. It is also worth noting that the shoulder becomes less pronounced as the temperature increases.
From the static structure factor it follows that the position of the first peak $k_m$ corresponds to the correlation length $r_m = 2\pi/k_m$. The quantity $r_m$ is related to the size of the atom. Comparing $r_m$ with the position $r_0$ of the first peak in the radial distribution function $g(r)$ and using the relation $r_m=2\pi/k_m$ ($k_m=2.22$~\AA$^{-1}$) we find that $r_m\simeq0.94~r_{0}$, where $r_0\simeq3.02$~\AA. Thus, the length of the average interatomic bond in liquid antimony at $T=923$~K is more than the effective diameter of the atom ($2.86$~\AA). It indicates the possible presence of metallic bonds (see Figure~\ref{gr_and_sk})~\cite{Vadrawaj_2022}.

\section{Identification of quasi-stable structures}

To identify ordered crystal-like structures in liquid antimony, the local orientational order parameters $q_4$ and $q_6$ were calculated~\cite{Steinhardt_1983_ql}. These order parameters are sensitive to structures with different types of crystal lattice symmetry, including rhombohedral, which is realised in the case of the Sb-I phase. The distributions of atoms over the orientational order parameters $q_4$ and $q_6$ were obtained using the expression:
\begin{equation}
	q_{l}(i) = \left( \frac{4\pi}{2l + 1} \sum\limits_{m=-l}^{l} \left| q_{lm}(i) \right|^{2} \right)^{1/2},\,\,\,l = \{4,\,6\},
\end{equation}
where
\begin{equation}
	q_{lm}(i)=\frac{1}{N_b(i)} \sum\limits_{j=1}^{N_b(i)}Y_{lm}(\theta_{ij},\,\phi_{ij}).
\end{equation}
Here $Y_{lm}(\theta_{ij},\,\phi_{ij})$ are the spherical harmonics, $\theta_{ij}$ and $\phi_{ij}$ are the polar and azimuthal angles, $N_b(i)$ is the number of nearest neighbours of the $i$-th particle. In the case of a close-packed bulk system consisting of atoms of the same type, it is sufficient to calculate the local order parameters $q_{4}$ and $q_{6}$. This makes it possible to identify all possible symmetry types of crystal lattices~\cite{Mokshin_PCCP_2017,Galimzyanov_AM_2019}.

\begin{figure}[ht!]
	\centering
	\includegraphics[width=1.0\linewidth]{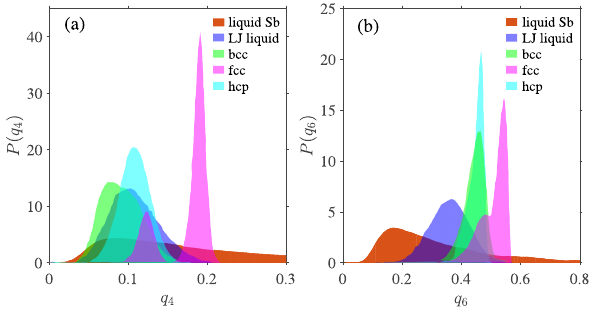}
	\caption{Distribution of atoms by the local orientational order parameters $q_4$ and $q_6$ obtained for liquid antimony (AIMD) and for Lennard-Jones liquid (LJ) as well as for the crystalline phases bcc, fcc and hcp.}
	\label{q4_q6}
\end{figure}

Figure~\ref{q4_q6} shows the obtained distributions for the order parameters $q_4$ and $q_6$. The absence of any crystalline phases in liquid antimony is evident. This can be seen by comparing the obtained results with data for the Lennard-Jones equilibrium liquid under similar thermodynamic conditions (i.e. $T/T_m\simeq1.05$)~\cite{Lechner_2008}. The Lennard-Jones liquid belongs to the class of simple liquids with a spherically symmetric type of atomic interaction such as in inert gases. Moreover, the shape and position of the distributions obtained for antimony melt are completely different from the distributions $P(q_4)$ and $P(q_6)$ typical for crystals with bcc, fcc and hcp lattices.

Thus, the obtained results indicate the absence of any crystalline structures in antimony melt at the considered temperature. However, short-lived structures may be present in the melt. To identify such formations, the neighbourhood times of pairs of atoms were calculated earlier at five different threshold values $R_c=3.4$, $3.8$, $4.2$, $4.6$ and $5.0$~\AA~\cite{Tsygankov_2023}. These values were taken to consider presence of bonded pairs of atoms. Their bond length must be more than the position of the main peak of g(r), but less than position of the shoulder edge $\approx 5.0$~\AA. The neighbourhood time $\tau(R_c)$ of two atoms is defined as the time during which these atoms are at a distance not exceeding the threshold value $R_c$~\cite{Mokshin_2020_gallium}. Figure 4 shows the distributions $P(\tau, R_c)$ of atomic neighbourhood times calculated for different values of $R_c$. These distributions characterise the neighbourhood time $\tau$ of atoms at a distance not exceeding $R_c$. It was shown that these distributions decay to zero at times longer than the structural relaxation time in the Williams-Landel-Ferry approximation $\sim1.6$~ps. Moreover, it has been shown from first-principles calculations that the structural relaxation time can be estimated by calculations of the intermediate scattering function of liquid antimony at the temperature $920$~K~\cite{Shi_2024_in_sn_sb}. This approach shows that the structural relaxation time is 0.5 ps. Thus, the structural relaxation time of liquid antimony can vary from $0.5$ to $1.6$ ps.
\begin{figure}[ht!]
	\centering
	\includegraphics[width=1.0\linewidth]{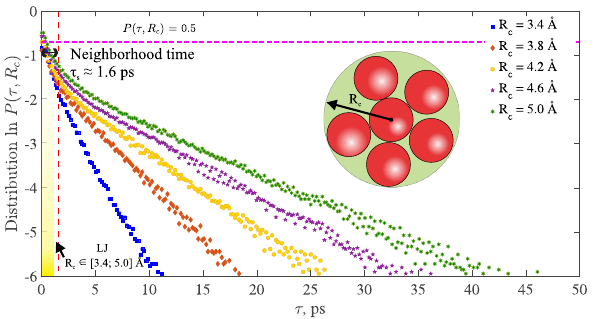}
	\caption{Distributions of neighbourhood times for a pair of atoms at different threshold values $R_c$. The red dashed line indicates the structural relaxation time calculated within the Williams-Landel-Ferry model and equal to $\tau_s = 1.6$\,ps. The yellow area shows the distributions by neighbourhood times for the Lennard-Jones liquid with $R_c\in[3.4;5.0]$~\AA, which rapidly decreases to zero on the picosecond time scale~\cite{Roos2013_structure_relaxation}.}
	\label{neigh_times}
\end{figure}

\begin{figure}[ht!]
	\centering
	\includegraphics[width=1.0\linewidth]{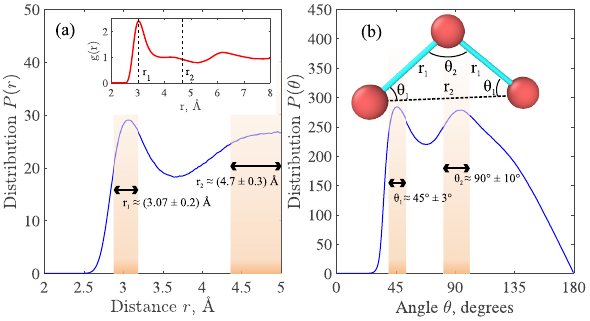}
	\caption{(a) Bond length distribution $P(r)$, which has two characteristic lengths $r_1\approx(3.07\pm0.2)$~\AA~and $r_2\approx(4.7\pm 0.3)$~\AA. The inset shows the radial distribution function $g(r)$, where the lengths $r_1$ and $r_2$ characterize the position of the first peak and shoulder, respectively. (b) Distribution over bond angles with characteristic angles $\theta_1$ $\approx$ $45^\circ$ $\pm$ $3^\circ$ and $\theta_2$ $\approx$ $90^\circ$ $\pm$ $10^\circ$. The inset shows a view of a triplet with the characteristic lengths and angles.}
	\label{bond_angle_distributions}
\end{figure}

As can be seen from Figure~\ref{neigh_times}, all distributions decay slower than the considered structural relaxation times. It shows that atoms in antimony melt form long-lived bonds. The distributions of bond lengths $P(r)$ for pairs of atoms and bond angles $P(\theta)$ of neighbouring atoms belonging to the long-lived quasi-stable structures were calculated. From the results shown in Figure~\ref{bond_angle_distributions} it can be seen that the bonds with lengths $r_1\simeq(3.07\pm0.2)$~\AA~and $r_2\simeq(4.7\pm0.3)$~\AA~are the most probable. These lengths coincide with the main maximum position (in the case of $r_1$) and shoulder position (in the case of $r_{2}$) of radial distribution function $g(r)$. The found distance $r_1$ is greater than twice the covalent radius of the antimony atom: $r_1\approx2r_{rad}$, where $r_{rad}=(1.45\pm0.05)$~\AA~\cite{Slater_1964}. This means that the atoms in the quasi-stable structures are bonded predominantly through the metallic type of bond~\cite{Yuntao_2018}. It should be noted that the metallic type bonds can be realised in liquid pnictogens as it is previously shown in Ref.~\cite{Kawakita_2018}.

Figure~\ref{bond_angle_distributions}(b) shows the distribution of bond angles for three neighbouring atoms, where the most probable angles are $\theta_1\simeq(45^{\circ}\pm5^{\circ})$ and $\theta_2\simeq(90^{ \circ}\pm15^{\circ})$. The presence of the angle $\theta_2\simeq90^{\circ}$ may be a consequence of the electron properties in the outer $6p$ orbital of pnictogens, where the bond angle is also $\sim90^\circ$~\cite{Reimers_2015_bond_angle}. Such characteristic angles are typical of all crystalline phases of antimony including the phase Sb-I, which is closest to the considered region of the phase diagram on the isobar $p=1$\,atm~\cite{Coleman_2018_sb_phase_diagram}. It follows that the characteristic angle $\theta_1$ is formed by the arrangement of atoms in the form of an isosceles triangle (triplet) with a right angle $\theta_2$ [see Figure~\ref{bond_angle_distributions}(b)]. The lengths of the edges of this triangle are determined by the characteristic lengths $r_{1}$ and $r_{2}$. 

\begin{figure}[ht!]
	\centering
	\includegraphics[width=1.0\linewidth]{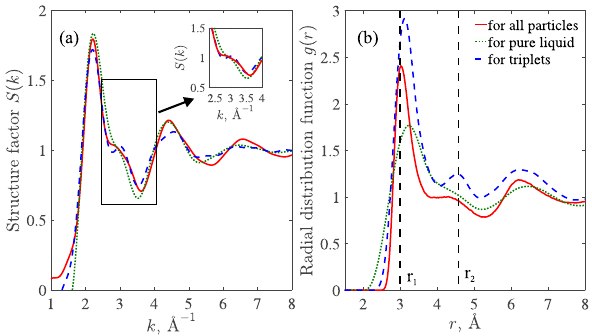}
	\caption{(a) Structure factor $S(k)$ and (b) radial distribution function $g(r)$ calculated separately for liquid atoms and separately for atoms forming triplets.}
	\label{separating_particles}
\end{figure}

\begin{figure}[h!]
	\centering
	\includegraphics[width=1.0\linewidth]{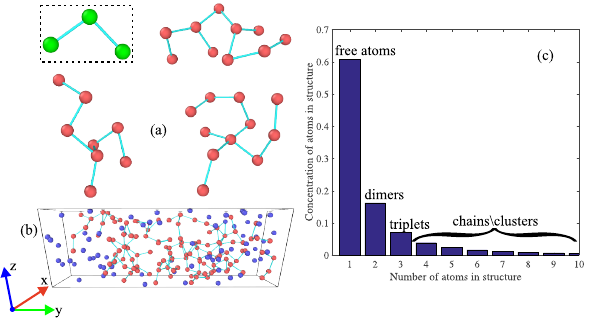}
	\caption{Three-dimensional chains formed by antimony atoms: (a) triplet and examples of chains with the average length of which is $\sim15$~\AA, (b) configuration of antimony melt with bonded atoms. (c) Concentration of quasi-stable structures as a function of the number of atoms in these structures.}
	\label{chains_examples}
\end{figure}

For a more detailed explanation of the shape of the static structure factor and the radial distribution function, the functions $S(k)$ and $g(r)$ were calculated separately for atoms of the liquid phase without triplets and separately only for atoms forming triplets. The found lengths $r_{1}$, $r_{2}$ and angles $\theta_1$, $\theta_2$ were used as criteria to search for these triplets. The results of Figure~\ref{separating_particles}(a) reveal that the shoulder in the function $S(k)$ almost completely disappears when calculations are performed for liquid, where the atoms forming triplets have been excluded from consideration. The shoulder is clearly visible in the function $S(k)$ obtained by considering only the triplet atoms. The shape of this function is close to the general static structure factor. Similarly, it is clear from Figure~\ref{separating_particles}(b) that the additional peak in the radial distribution function $g(r)$ is practically absent in the case of the melt without triplets~\cite{Galimzyanov_Mokshin_2017,Mokshin_TMF_2021}. Therefore, we can conclude that the presence of the shoulder and additional peak in the functions $S(k)$ and $g(r)$ can be explained by the presence of quasi-stable structures formed by triplets. Thus, we define triplets as localized particles with the bond angles $45^\circ\pm3^\circ$ and $90^\circ\pm10^\circ$ as well as the bond lengths, which can vary in the interval $[2.88; 3.27]$~\AA\, and $[4.0; 5.0]$\,\AA. If the nearest neighbouring particles, including those in possible dimers, contribute to the formation of the main maximum of the g(r) function, then the appearance of the shoulder in g(r) is a result of pair correlations at the characteristic distance $\approx 4.7$~\AA, which is typical for the detected triplets (see Figure 6b in the manuscript).

The results of the cluster analysis show that there are two forms of triplets: (i) single triplets, Sb-Sb-Sb; (ii) triplets that are elementary structural units capable of forming branched (percolating) structures in the form of chains, -Sb-Sb-Sb- [see Figure~\ref{chains_examples}(a)]. The configuration of these chains changes due to the thermal movement of the atoms and the rearrangement of the triplets. At the same time, the characteristic lengths and angles between the bonded atoms are saved, giving the effect of the presence of long-lived structures in antimony melt. Under the considered thermodynamic conditions, the fraction of atoms involved in the formation of these chains can be up to half of the atoms in the system, as can be seen in Figure~\ref{chains_examples}(b). It is remarkable that similar results were obtained before for liquid bismuth, where the presence of chains was discovered. The elementary units of these chains are also triplets~\cite{Galimzyanov_2023_liquid_bi}. The similarity of these results can be associated with a common principle of organisation of the liquid structure near the melting temperature. 

In a first approximation, it may appear that the elementary unit of quasi-stable structures is a dimer. However, the concept of bond angle is not applicable to dimers because the angle cannot be clearly distinguished. Due to this fact, we have to consider such a structure as elementary, which has a minimal size and carries all information about the geometrical characteristics of quasi-stable structures according to the obtained distributions of bond lengths and angles. The triplet is the elementary structure that fulfills these requirements.

The concentration of quasi-stable structures in the form of free atoms (Sb), dimers (Sb-Sb), triplets (Sb-Sb-Sb) and formations of four or more atoms was determined. Figure~\ref{chains_examples}(c) below shows that the concentration of free atoms is highest. The obtained result shows that the structure of liquid antimony is a set of free and bound atoms, which form the observed quasi-stable structures. Also, total fraction of all bound atoms is 39\%. 
Similar results was obtained by H. Tanaka, where structure of liquid antimony was considered as set of free atoms and bound structures which are created and annihilated spontaneously~\cite{Tanaka_2020}. It was shown that total fraction of bound atoms in liquids with an anomalous structure can be up to half of all atoms.

Quasi-stable structures of $4$ or more atoms are formed by combining triplets, dimers and free atoms. Compared to dimers and triplets, such structures have a much lower concentration. Thus, despite the predominance of dimers with characteristic lengths of $3.07$~\AA~and $4.7$~\AA, we consider triplets to be the basic elementary unit of quasi-stable structures, with characteristic angles of $\simeq45^{\circ}$ and $\simeq90^{\circ}$ in addition to the above characteristic lengths. It has been demonstrated in~\cite{Shor_2012} that two characteristic bond lengths are required to describe the anomaly in the liquid antimony.

In addition, the crystallization of supercooled liquid antimony was analysed in article~\cite{Ropo_2017}. It was shown that during crystallization the formation of $\simeq90^{\circ}$ bond angle structures is favored. This result indicates the existence of a general principle of structure formation during the crystallization process, which allows us to introduce the concept of "elementary unit" of quasi-stable structures. The smallest structure with this property is the detected triplet, which turns out to be quasi-stable at the considered temperature $T = 923$ K.

\section{Energy state of quasi-stable structures}
\begin{figure}[h!]
	\centering
	\includegraphics[width=0.6\linewidth]{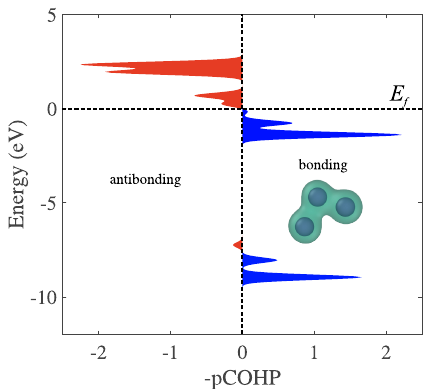}
	\caption{Averaged pCOHP for all atoms of a single triplet with a representation of its isoenergetic surface. The energy contributions tending to break interatomic bonds are indicated by red filling, while the energy contributions tending to bond atoms together are coloured blue. All energy values are considered relative to the Fermi energy $E_f$ shifted to zero.}
	\label{cohp}
\end{figure}

Information about the energy state of quasi-stable formations has been obtained using the Crystal Orbital Hamilton Population (pCOHP) method (see Fig.~\ref{cohp}) \cite{Deringer_2011_pcohp,Maintz_2016_cohp}. This method is based on the analysis of contributions of atomic orbitals to the interaction of atoms with each other using projective density matrices, which allows one to extract useful information about the interaction of atoms directly from ab-initio simulation. In particular, the bonding energy of the atoms in the triplet structure can be estimated indirectly from the integral over the pCOHP curve up to the Fermi energy $E_f$~\cite{Janine_2022_cohp,Dronskowski_1993_cohp}. The more stable the bond between the atoms, the higher the value of the pCOHP integral. For example, it has been shown by pCOHP calculations that in bound structures the value of the pCOHP integral is negative, as in the case of crystalline tellurium~\cite{Decker_2002_cohp_te}, which in the liquid state has the same structural anomaly as antimony melt~\cite{Akola_2010_liquid_te}.

In the present work, the following conditions were used to obtain the necessary data before the pCOHP calculation: ab-initio calculations were performed using the Projected Augmented Wave (PAW) method and the Perdew-Burke-Ernzerhof (PBE) approximation with a kinetic cutoff energy of $400$~eV~\cite{Perdew_1996_pawpbe}. The size of the k-point grid was $12\times12\times12$. The pCOHP integral of the triplet state energy up to the Fermi level is $E_t\simeq-1.8$~eV. The negative value of the pCOHP integral indicates that the triplet structure may be stable (see Figure~\ref{cohp}). On the other hand, the average kinetic energy of atoms forming triplets is $\simeq5\cdot10^{-4}$~eV at the temperature $T=923$\,K, which is lower than the average kinetic energy of unbound atoms, which is $\simeq2.7\cdot10^{-3}$~eV. This confirms quasi-stability of triplet structures formed in the antimony melt near melting point.

\section{Conclusions}

Analysis of the local structure of liquid antimony near its melting temperature reveals the existence of quasi-stable structures in the form of triplets. The results also reveal that these triplets can form chains of different lengths, which can span the entire system. The found neighbourhood time distributions show that a characteristic lifetime of these triplets and the chains can be about tens of picoseconds, which is an order of magnitude longer than the structural relaxation time of the melt. The quasi-stability of these triplets is also confirmed by the negative energy value of the pCOHP integral as well as by calculated small value of the kinetic energy relative to the unbonded atoms. However, liquid antimony is not the only example of a unique pnictogen in which structural anomalies are observed. The quasi-stable structures formed by triplets can be found in liquid pnictogens near the melting temperature when moving along the group towards light atoms such as arsenic and phosphorus~\cite{Mayo_2013_trends,Yang_2021_liquid_phosphorus}. For example, the metallic behaviour of liquid arsenic was discovered, indirectly confirming the existence of quasi-long-lived structures~\cite{Li_1990_liquid_arsenic}. Moreover, shoulders and peaks in the structure factor are also observed in other monatomic melts. Here it is worth highlighting the work of Hafner et al. which provides a detailed analysis of experimental data on the structure of high density liquid polyvalent melts from Groups I-VI of the Periodic Table~\cite{Hafner_1992}. Such polyvalent melts include mercury, gallium, indium, silicon, germanium and others, where additional peaks and shoulders are also observed in the structure factor. However, in the case of liquid gallium~\cite{Mokshin_2020_gallium}, the absence of any ordered structures has been shown, making the task of deciphering the unusual structures of monatomic liquids non-trivial.

\appendix
\section*{Appendix: Simulation details and applied methods}
{\it Simulation procedure}. -- A first-principles (\textit{ab-initio}) molecular dynamics simulation of liquid antimony has been carried out using the interatomic interaction calculated by the Born-Oppenheimer method with the ultrasoft exchange-correlation functional~\cite{Kresse_1993_1,Kresse_1996_1,Kresse_1996_2,Kresse_1994}. The thermodynamic state with the temperature $T=923$~K at the pressure $p=1$~atm was considered. The temperature was controlled by the Nose-Hoover thermostat. The system has the constant quantitative density $\rho_n = 0.0320$~\AA$^{-3}$, which corresponds to the experimental value at the temperature $T=923$~K~\cite{Waseda_1971_experimental_data}. Atoms in quantity of $384$ were placed inside a rhombohedral simulation cell with side lengths $L_x = 17.54$~\AA, $L_y = 17.54$~\AA, $L_z = 45$~\AA~\cite{Lejaeghere_2014_sb_struct,Vereshchagin_1965_sb_struct,Akselrud_2003_sb_struct}. The considered time step is $\Delta t=0.003$~ps.

{\it Structure analysis}. -- The radial distribution function $g(r)$ was calculated based on molecular dynamics simulation data using expression~\cite{Khusnutdinoff_2018_gr_sk}:
\begin{equation}
	g(r) = \frac{V}{4\pi r^2 N}\left\langle\frac{\Delta n(r)}{\Delta r} \right\rangle,
\end{equation}
where $V$ is the volume of the system, $N$ is the number of atoms in the system, $\Delta n(r)$ is the number of atoms located in a spherical layer of thickness $\Delta r$, $r$ is the distance between two atoms. The static structure factor was determined based on the calculated function $g(r)$ through the Fourier transform of the following form:
\begin{equation}
	S(k) = 1 + 4\pi\rho_n\int\limits_{0}^{\infty}r^2\left[g(r) - 1\right]\frac{\sin(kr)}{kr}dr,
\end{equation}
where $k$ is the wave number.

\section*{Acknowledgements}
This work is supported by the Kazan Federal University Strategic Academic Leadership Program (PRIORITY-2030).

\end{document}